\documentclass[conference,flushend]{iaria} 
\usepackage{nopageno}
\usepackage[backend=biber]{biblatex}
\usepackage{graphicx}
\usepackage{fancyhdr}
\usepackage[export]{adjustbox}
\usepackage{array}
\usepackage{helvet}

\newcommand{\figref}{Figure~\ref}

\title{Hip Fracture Patient Pathways and Agent-based Modelling}
\author{
  \IEEEauthorblockN{%
    Alison N. O'Connor,\orcidlink{0000-0002-8331-2901}, Stephen E. Ryan,\orcidlink{0000-0002-9704-7248}, Gauri Vaidya,\orcidlink{0000-0002-9699-522X}, Paul Harford, and Meghana Kshirsagar,\orcidlink{0000-0002-8182-2465}}
  \IEEEauthorblockA{%
    Department of Computer Science\\
    University of Limerick\\
    Limerick, Ireland\\
    e-mail: {\tt$\lbrace$alison.oconnor\,|\,stephen.e.ryan\,|\,gauri.vaidya\,|\,meghana.kshirsagar$\rbrace$@ul.ie}
    }%
    {\tt Paul Harford | 23190973@studentmail.ul.ie}
    }
   
\begin{document}
\maketitle
\begin{abstract}
Increased healthcare demand due to ageing populations is significantly straining European services. 
In Ireland patients presenting in Emergency Departments with hip-fracture injuries should be admitted to Orthopaedic wards or Theatre within four hours.
However, the Irish Hip Fracture Database reports that less than 25~\% of patients experience this pathway.
Digital solutions, including advanced modelling techniques, offer a promising solution to optimising patient flow without impacting day-to-day healthcare provision. 
In this work, we outline ongoing research that aims to optimise healthcare provision for hip-fracture patients through improved data integration and agent-based simulations.
We demonstrate how this technology, through enhanced connectivity, can improve healthcare service for hip-fracture patients.
\end{abstract}

\pagestyle{empty}

\section{Introduction}\label{sec:intro}

Several key factors contribute to recurring overcrowding in Irish hospitals, many of which reflect deep-rooted systemic issues. 
A major issue driving increased healthcare demand, is the complex medical needs and increased Emergency Department (ED) attendance frequency required for older patients (patients aged $\geq 65$ years).
Ireland's rapidly ageing population directly correlates to increased patient demand~\cite{WALSH2021}, and the European Union (EU) notes a similar trend~\cite{EUROSTAT2020,EC2018}.
Data from 2022 paints a stark picture of the severity of the problem. 
In 2022, over 70\% of Irish EDs exceeded their maximum capacity, leaving staff and resources perilously stretched~\cite{HIQA2022c}. 
Even more concerning, 80\% of these hospitals reported significant staffing shortages, further compounding the difficulties in managing patient care effectively~\cite{CULLEN2022, POWER2022a}.
The shortage of trained healthcare workers has been a persistent issue across the EU ~\cite{LOOI2024}. 
We're now faced with healthcare environments increasingly unable to meet demand~\cite{HSE2022}.

In response to healthcare challenges, Ireland's regulatory body issued a series of recommendations emphasising the importance of effective workforce planning~\cite{HIQA2022b}.
Digital technologies offer a promising solution to the challenge of resource optimisation in dynamic environments, however, these efforts must be paired with forward-thinking strategies to prepare for future challenges. 
More recent literature has demonstrated a strong appetite within Ireland for healthcare digital technologies~\cite{WHELAN2024}.
While Ireland has committed to digitising healthcare the Irish health ecosystem lags far behind European counterparts~\cite{DOH2024a}, making it difficult to ascertain what can be done with existing healthcare datasets.
Without decisive action on how we can best use current digital records, the overcrowding crisis will persist, with dire societal consequences.
The work focuses on how existing Irish healthcare data can be used to optimise resources.

Resource optimisation is not a new concept, for example, `Lean manufacturing' is a systems engineering approach aimed at increasing operational efficiency through waste minimisation.
It has been highly successful in manufacturing environments~\cite{VIGNESH2016} and has been adapted to other areas like supply chain management~\cite{CHEN2013} and healthcare~\cite{RUTMAN2015}.
Lean application is process specific and its successful implementation is strongly linked with leadership and company culture~\cite{TORTORELLA2021}.
The case-by-case implementation to different hospitals, with different resources, and indeed different cultures, means that it is almost impossible to apply a specific lean method across all environments.
These challenges mean that Lean processes are, generally, applied to specific (but often reoccurring) sub-processes. 
Hence rather than the large-scale full process implementation we see in manufacturing environments, lean processes in healthcare are not only limited in scope but also cannot be generalised to other locations.
These limitations are exacerbated by the fact that lean systems tend to perform poorly when deployed to dynamic processes~\cite{AZADEGAN2013}.

We propose that Artificial Intelligence (AI) and Machine Learning (ML) can be used to leverage existing data beyond its original collection purposes.
These tools, which have proven abilities to resolve highly non-linear (i.e., dynamic) relationships, can be deployed to bypass the limitations of Lean methods~\cite{SHAHIN2024}.
The goal of this research is to improve patient flow for hip-fracture patients through resource optimisation via agent-based modelling.
The increased understanding of resource requirements will enable healthcare staff to identify and propose targeted interventions that can be deployed in the real-world. 
Targeted interventions, deployed in an incremental fashion, can reduce disruption within the acute care setting while also providing adequate feedback on algorithmic performance.

In Section~\ref{sec:ai} we introduce Artificial Intelligence (AI) in the context of healthcare services and address why these tools are considered critical to future healthcare operational planning. 
We also briefly discuss limiting factors surrounding the use of AI tools in healthcare.
Section~\ref{sec:hipfracture} describes a framework of AI tools that can be used to improve patient flow for hip-fracture patients. 
The potential impacts of this research are addressed in Section~\ref{sec:impact}, with concluding remarks and future work outlined in Section~\ref{sec:conclusion}.

\section{Artificial Intelligence (AI)}\label{sec:ai}
\begin{figure}[b]
    \centering
    \includegraphics[width=\linewidth]{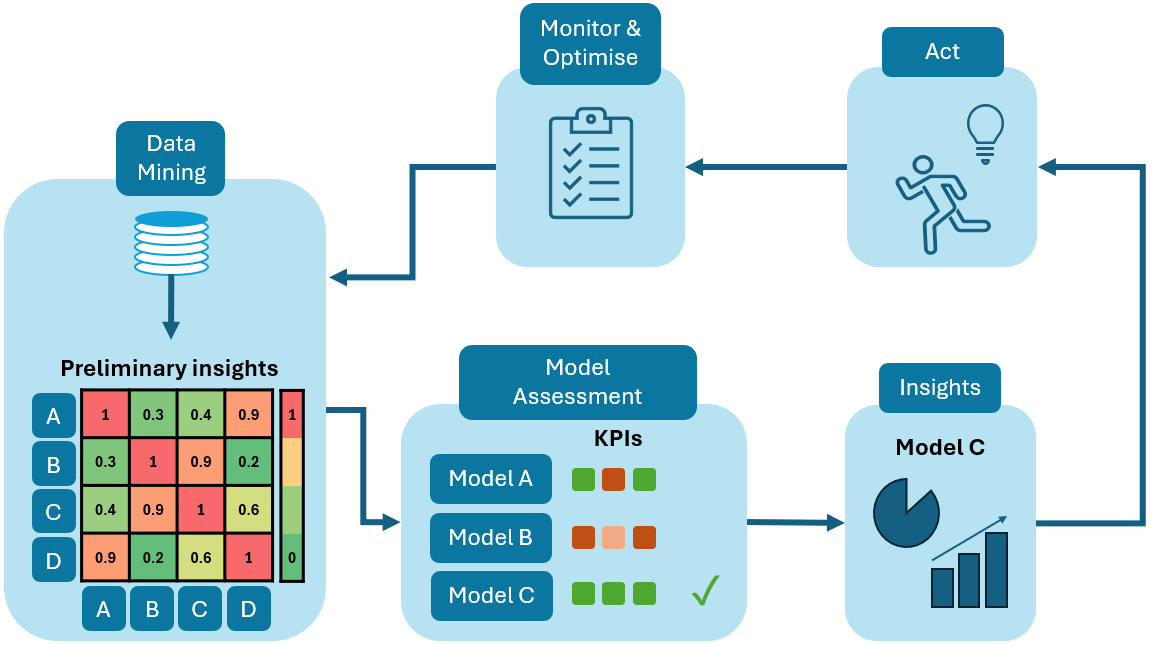}
    \caption{Infographic outlining machine learning model selection and assessment of actionable insights.}
    \label{fig:data_mining}
\end{figure}

Digital technology is revolutionising healthcare by offering new insights through data analysis.
Automation reduces the need for manual coordination between departments, and reduces the administrative burden of healthcare workers~\cite{ABBING2016}, enabling them to focus on patient care, thus increasing the overall efficiency of the hospital~\cite{KELLY2022}.
ML is a sub-field of AI that addresses the methods used by AI to perform specific tasks~\cite{EP2022}.
ML algorithms have demonstrated abilities in predictive analytics and resource optimisation which can be used to improve operational efficiency~\cite{RATHORE2021b}.
One of the most significant benefits of ML for healthcare applications is its ability to leverage existing data beyond the original data collection purposes.
In \figref{fig:data_mining}, we illustrate how existing data can be assessed to identify appropriate ML models providing actionable insights that can be used to optimise workflows.
The integration of AI/ML has long-term financial benefits for healthcare systems. 
One example is fewer instances of under-staffing, avoiding the financial strain associated with agency based staff, which cost Ireland in excess of €640 million in 2023 alone~\cite{WHITE2024}. 

While digital solutions offer the potential for transformative change in healthcare provision, it is not without limitations.
One limitation is the lack of transparency in algorithm outputs.
Simulation methods that perform well on diverse datasets are generally so complex that it is impossible for end-users to interpret model outputs~\cite{YANG2022b}. 
Enhancing explainability for algorithm outputs remains an active research area~\cite{ALI2023b}.
Another limitation is the requirement for large amounts of sensitive personal information which is highly restricted by Europe's General Data Protection Regulation (GDPR).
GDPR has significantly affected researchers' abilities to access meaningful datasets to build, test and/or validate AI/ML models~\cite{NEPELSKI2021a}.
Despite these limitations, the immense societal benefits of digital health  has been recognised at a European level~\cite{EC2021}.
For hospitals already facing strain due to rising patient numbers and limited resources, these technologies are fast becoming a critical requirement for ensuring that patient care is delivered efficiently without compromising quality. 
The benefits extend to long-term cost reductions, better resource management, and improvements in patient outcomes, all of which contribute to a more sustainable and effective healthcare system~\cite{BUIJS2024}.

The question is no longer whether we will implement these technologies, but rather how we can do so effectively and responsibly. 
In this ongoing collaborative research project, we are investigating the integration of ML technologies into a hospital in the mid-west of Ireland. 
This initiative aims to explore applications of ML in areas such as predictive patient demand, patient flow optimisation, and resource management, potentially revolutionising healthcare delivery through connected applications.
Through this research, we seek not only to enhance patient care and hospital efficiency at a local level but also to establish a framework that could scale nationally, transforming the broader healthcare landscape. 
Ultimately, the project aspires to position Ireland at the forefront of healthcare innovation, with the potential for these technologies to be adopted on a global scale.

\section{Hip fractures: a case study}\label{sec:hipfracture}
The ageing population of Europe has been identified as a significant contributor to overcrowding.
Considering the correlation between age and fragility it is not surprising that older people (+65) are at a higher risk of experiencing complex medical fractures~\cite{WALSH2021}.  
\textcite{KANIS2021} indicates the European cost of fracture services to be in excess of €56 billion (2019) with €290 million attributed to the Irish exchequer.
Considering patient flow for a specific sub-set of patients, in this case hip fracture patients, we aim to identify specific causes of poor patient flow enabling targeted resource allocation that impacts both subset and larger population groups.  
Here we detail our ongoing research into how digital technologies can be used to improve system operations across an acute care setting for this patient group. 

Mid-west Ireland has a population of approximately 500,000 with healthcare provided by the UL Hospital Group (ULHG).
The ULHG comprises six facilities with University Hospital Limerick (UHL) operating as the centralised location for critical care~\cite{HSE2023b}.
UHL monitor patients from admission in the ED through the wider acute care setting (e.g., surgical and/or medical procedures) up to patient discharge.
Hip fracture patients in the Mid-west requiring longer-term care but with reduced medical needs are often facilitated by other facilities in the UHLG.
UHL liaise with the National Office of Clinical Audit (NOCA).
NOCA maintains the Irish Hip Fracture Database (IHFD) which collates hip-fracture patient information nationally. 
The purpose of the IHFD is to assess individual hospital compliance against seven key performance indicators (KPIs) including patient time to ward, patient time to the operating theatre~\cite{KELLY2022}.
While the overall purpose of the IHFD is to assure quality of healthcare provision, the richness of the data within the IHFD can be exploited far beyond this task.  
Forecasts predict that hip fracture hospitalisations in Ireland could increase threefold by 2046~\cite{KELLY2018}, which would have a significant impact on HSE resourcing.
In 2019 the +65 cohort increased by 1.2\% while a 4.2\% increase in hip fracture cases was reported~\cite{JONES2020}.

\begin{figure}
    \centering
    \includegraphics[width=\linewidth]{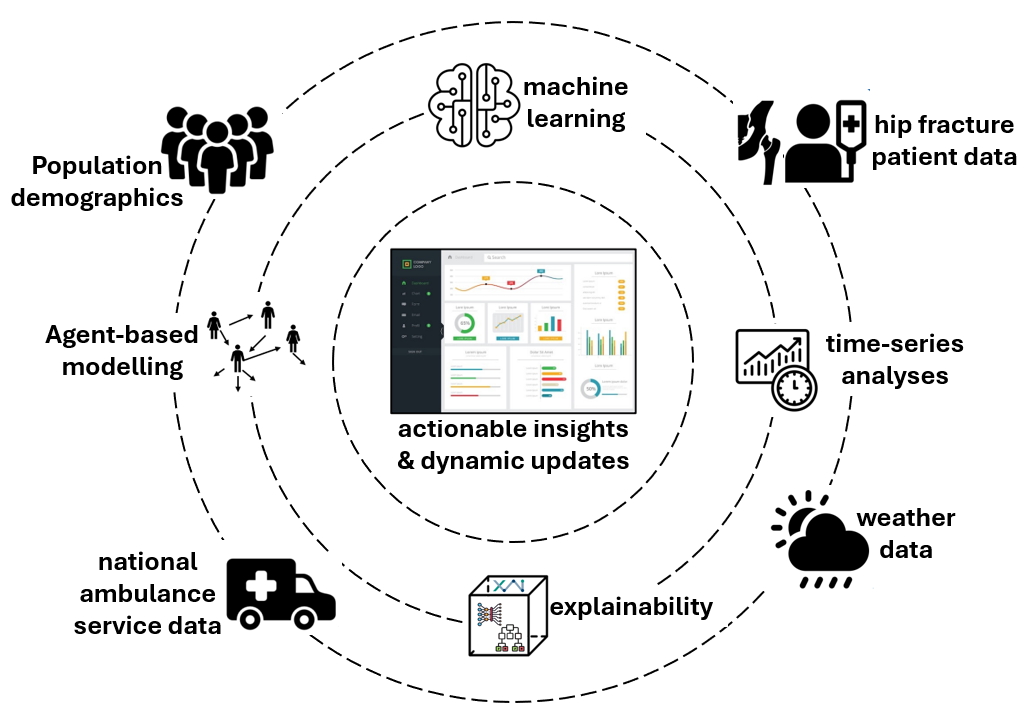}
    \caption{Pipeline of tools investigated in this research.}
    \label{fig:pipeline}
\end{figure}

In \figref{fig:pipeline}, we illustrate the specific datasets and tools of interest to this work.
The outer circle of \figref{fig:pipeline} shows datasets that can be used to provide additional context of value and relevance to hip-fracture patients. 
Immigration, due to global challenges, has significantly increased Ireland's population in recent years, with the population increasing by over 98,000 in the year ending April 2024~\cite{CSO2024a} alone.
The share of elderly ($\geq 65$ years) immigrants has grown in the EU to account for approximately 21\% of the immigrant population, and approximately 6\% are over 75 years~\cite{OECD2023}.
This has obvious implications for age-related illnesses such as hip fractures, where the mean age of fractures is 83 and 84 years for men and women, respectively.
Analysing national census databases~\cite{CSO2024b} enables us to statistically interrogate how national and regional population demographics may affect service demand.
The IHFD will provide contextual information related to the average patient age, and the frequency of hip-fractures both nationally and regionally.

Most hip-fracture patients are posited to arrive via ambulance services but, to our knowledge, this relationship has not been confirmed analytically. 
Using National Ambulance Service (NAS) records to assess the frequency of NAS requests will enable us to better understand the correlation between this patient group and NAS resource demand. 
Similarly previous \textcite{STANLEY2023} indicated an association between national weather warnings and hip fractures.
However, national weather warnings frequently do not reflect regional weather and the specific type of weather (e.g., rain, ice, snow etc.) was not addressed.
\textcite{YEUNG2020} however demonstrates a strong correlation between night-freezing weather events and fall-related injuries.

The middle circle of \figref{fig:pipeline} shows the machine learning algorithms identified as potential tools for analysing data.
Time-series analysis will be employed to predict seasonal hip-fracture service needs, the average patient length of stay, and the average time spent in EDs and/or surgical units.
We previously (Section~\ref{sec:intro}) outlined how explainability in AI/ML healthcare is a core requirement for European healthcare services. 
In this research we will investigate multiple types of ML algorithms (see \figref{fig:data_mining}) and assess their performance across an array of KPIs.
One of the KPIs in our research will centre around explainable artificial intelligence comparing so-called `black-box' algorithm performance with more interpretable `white- or glass- box' methods.
More advanced simulation methods such as agent-based models, which simulate interaction events, can be used to not only identify the most common patient pathways but also indicate pathway bottlenecks. 
Agent-based models enable us to interrogate multiple `what-if' scenarios without impacting day-to-day functioning of acute environments.
Test cases of interest to this work include scenarios such as:
\begin{enumerate}
    \item What is the impact of ageing on current service demand, and how will this change in the future based on population changes?
    \item What are the optimal staffing resources required to maximise ED service?
    \item How do optimal resource requirements change for specific patient groups?
    \item What is the average length of stay for hip-fracture patients in UHL and how does this compare with other facilities in the UHLG?
\end{enumerate}

The inner circle of \figref{fig:pipeline} illustrates our expected end goal.
Our suite of tools will be used to create a dynamic dashboard highlighting, in an easily interpretable manner, key information related to this patient subgroup.
As this work is ongoing the dashboard display has not yet been fully conceptualised but is likely to include information such as: current number of patients, model prediction of future patients, and the average patient wait time in the ED.

\section{Impact} \label{sec:impact}

In the context of the IHFD, it is interesting to note that data collection frequently occurs post-intervention which limits the ability of healthcare workers to effect change in a timely manner.
Using digital platforms like that illustrated in \figref{fig:pipeline} to automate both data collection and data analysis, open the possibility of real-time information.
quicker transfer times to surgery for hip fracture patients can reduce the risk of complications such as infections or blood clots, which would otherwise require additional medical interventions and extended hospital stays~\cite{KELLY2022}.
Further, this type of system can be integrated with existing hospital data collection systems reducing the need for manual, repetitive administration.
Real-time data access can reduce treatment delays directly benefiting patients.
Accurate ML models can support operational decision-making around workforce planning and resource allocation.
Armed with model predictions, hospitals can preemptively adjust staffing levels, prepare operating theatres, and ensure that enough beds are available to meet patient demand. 
This type of foresight allows hospitals to avoid being overwhelmed, optimises resource use, and prevents operational inefficiencies.
By better managing both human and material resources, hospitals can operate more cost-effectively, benefiting not only the institution but also the broader healthcare system.
Over time, these improvements contribute to significant cost reductions by improving hospital efficiency and reducing the need for expensive follow-up treatments due to delayed care.
Additionally, better resource management means fewer instances of over- or under-staffing, optimizing the use of available personnel and avoiding the financial strain of overtime pay or temporary staff hires during peak periods.
The integration of datasets from diverse institutions enhances communication between multiple parties which can provide an improved patient experience and, more importantly, improved patient outcomes.
This research can act as a starting point for larger interoperable systems extending beyond acute care environments to incorporate national and regional information, as outlined here.
Future work includes aims to integrate acute- and community- care (e.g., General Practitioners) systems providing a more holistic healthcare system for Irish citizens.

\section{Conclusion and Future Work}\label{sec:conclusion}
In summary, the integration of digital technology, AI, and ML in healthcare represents a transformative opportunity to address the pressing challenges faced by emergency departments, particularly in the context of the IHFD.
The continuous surges in patient volume, especially in emergency settings, have led to significant overcrowding and strain on the healthcare system. 
By leveraging data-driven insights, hospitals can optimise resource allocation, improve patient flow, and enhance compliance with KPIs.
The methods discussed here offer a unique insight into how acute care data collection in Ireland can be leveraged beyond its original purpose providing actionable insights that can radically improve workflows without negatively impacting daily acute care operations.
These technologies not only facilitate the forecasting of patient admissions, but also enhance the prioritisation of care for high-risk patients, ultimately leading to improved patient outcomes. 
Moreover, the application of AI and ML allows for the continuous optimisation of patient care pathways, ensuring that healthcare resources are used effectively and efficiently.

One significant challenge for this research relates to GDPR restrictions and dataset access.
The ethical requirements for healthcare related projects in Ireland are demanding, often requiring researchers in third level institutions to have support from medical staff operating within Ireland's public healthcare system.
This can be problematic for researchers in fields like computer science that may not have network connections in medical fields.
Ethical applications are generally submitted a month in advance of the committee hearing and applicants may be invited for in-person review.
While committees meet monthly there are limitations to how many applications can be reviewed at any given time.
Even with medical staff support and ethical approval, dataset access is not guaranteed as additional legal requirements around data sharing must be arranged.
Most research projects have defined funding timelines and the slow speed of dataset access presents significant challenges in terms of research outputs.
To mitigate these risks researchers may request anonymised and/or aggregated dataset access.
While this eases the ethical application processes it can have significant implications on ML algorithm precision, especially in terms of prediction capability.
Compounding these challenges is the additional limitations on integrating multiple datasets - which could potentially lead to personal data recovery.
Together these limitations not only act to reduce algorithm performance, but also severely limit the potential to uncover relationships between acute- and community-care regimes.
While implementing AI/ML technology on restricted and reduced datasets is not ideal it does offer us the opportunity to identify areas for further exploitation.
Preliminary results, generated from these reduced datasets, can be used to showcase the potential of these technologies to improve our healthcare system which may garner further support for future research works.
This project represents an opportunity to test the precision and reliability of explainable AI/ML methods such as:
\begin{itemize}
    \item Predicting next weeks/months/years' hip fracture cases based on historical accounts and population growth/contraction.
    \item Classification of patients based on risk profile and treatment requirements (e.g., prioritising high risk patients).
    \item Clustering analyses to identify patterns in patient movement and resource use to highlight bottlenecks.
    \item Further patient pathway optimisation through reinforcement learning.
\end{itemize}
Only by testing these approaches we can define the limitations and potential uses of these algorithms in supporting healthcare workers.

One future area of research of specific interest is the potential for Radio-Frequency IDentification (RFID) to trace patient pathways from ED arrival through the acute care setting.
RFID technology uses radio waves to wirelessly identify object location and is commonly used in postal services to track packages.
This technology offers exceptional granular level information about the experienced patient pathway, which is not currently captured by existing healthcare systems.
Upon admission, patients with RFID bracelets transmit signals to strategically placed sensors throughout the hospital.
These sensors log patient locations and movements, delivering real-time data on each patients whereabouts within the facility.
Analyzing the data from RFID sensors allows hospitals to identify the most efficient pathways patients take during their hospital stay.
This information can be used to reorganise hospital layouts and workflows, ensuring that patients move smoothly through necessary departments while minimizing delays.
Moreover, understanding patient movement patterns enables better resource management, ensuring that staff and equipment are available where and when they are most needed.
For instance, if a specific surgical ward consistently experiences delays in patient transfers, this data can prompt an investigation into staffing or logistical issues that may be causing inefficiencies.
To effectively communicate findings and insights, data visualization tools can be employed to create dashboards and reports summarizing key metrics and trends.
These visualizations help hospital administrators quickly identify areas needing improvement and support data-driven decision-making.
By employing these methods, hospitals can harness the power of AI and ML, along with RFID technology, to optimize patient flow, improve compliance with KPIs, and ultimately enhance patient outcomes.
This comprehensive approach to data analysis and resource management is essential for addressing the challenges faced by emergency departments and ensuring efficient healthcare delivery in Ireland.

As healthcare systems strive to improve efficiency and patient care, the adoption of these digital technologies and analytical methods will be crucial. 
The potential for long-term cost savings, improved patient outcomes, and more effective resource management underscores the importance of embracing this technological evolution in healthcare. 
Ultimately, by harnessing the power of ML hospitals can not only meet current demands but also proactively prepare for future challenges, creating a more resilient and responsive healthcare system for all.

\section*{Acknowledgements}
We gratefully acknowledge the expertise, guidance and support of our UHL colleagues Damien Ryan, Carrie Garavan and Fergal Cummins. 
This project is funded under the National Challenge Fund which was established under the government’s National Recovery and Resilience Plan (NRRP), funded by the EU’s Recovery and Resilience Facility. The fund is coordinated and administered by Taighde Éireann – Research Ireland grant 22/NCF/DR/11289.

\printbibliography
\end{document}